# NRSSPrioritize: Associating Protein Complex and Disease Similarity Information to Prioritize Disease Candidate Genes

Razieh Abdollahi[a], Sama Goliaei[b], Zahra Razaghi-Moghadam[a,c*] and Morteza Ebrahimi[a]

[a] *Faculty of New Sciences and Technology, University of Tehran, Tehran, Iran.*
[b] *University of Tehran, Tehran, Iran.*
[c] *School of Biological Sciences, Institute for Research in Foundation Sciences (IPM), Tehran, Iran.*
[*] *Corresponding author: Dr. Zahra Razaghi-Moghadam, email:razzaghi@ut.ac.ir*

{*s.r_abdollahi, sgoliaei, razzaghi, mo.ebrahimi}@ut.ac.ir*

***Abstract:*** The identification of disease-associated genes has recently gathered much attention for uncovering disease complex mechanisms that could lead to new insights into the treatment of diseases. For exploring disease-susceptible genes, not only experimental approaches such as genome-wide association studies (GWAS) have been used, but also computational methods. Since experimental approaches are both time-consuming and expensive, numerous studies have utilized computational techniques to explore disease genes. These methods use various biological data sources and known disease genes to prioritize disease candidate genes. In this paper, we propose a gene prioritization method (NRSSPrioritize), which benefits from both local and global measures of a protein-protein interaction (PPI) network and also from disease similarity knowledge to suggest candidate genes for colorectal cancer (CRC) susceptibility. Network Propagation, Random Walk with Restart, and Shortest Paths are three network analysis tools that are applied to a PPI network for the purpose of scoring candidate genes. Also, by looking through diseases with similar symptoms to CRC and obtaining their causing genes, candidate genes are scored in a different way. Finally, to integrate these four different scoring schemes, Technique for Order Preference by Similarity to Ideal Solution (TOPSIS) and Analytic Network Process (ANP) methods are applied to obtain appropriate weights for the above four quantified measures and the weighted summation of these measures are used to calculate the final score of each candidate gene.

NRSSPrioritize was validated by cross-validation analysis and its results were compared with other prioritization tools, which gave the best performance when using our proposed method.

*Keywords:* Gene prioritization, Protein-protein interaction network, Symptoms, Colorectal cancer.

## 1. Introduction

One of the most important challenges in disease treatment is the identification of causing genes that help us to design medical protocols. Genome-Wide Association Studies (GWAS) are new techniques for the identification of chromosomal intervals containing suspected disease related genes. These studies search for the genomes of single nucleotide polymorphisms (SNPs) that are not rare. However GWAS studies are not very accurate for detecting the exact gene related to a disease because they suggest hundreds or thousands of genes. Investigating all the genes suggested by GWAS in order to find the desired gene using experimental methods is very expensive and time consuming. Computational methods prioritize these genes and help us to focus on a smaller set of genes. For this purpose, computational methods utilize disease genes that are suggested experimentally as seed genes and prognosticate candidate genes for further studies. They use various data to solve the problem and some of them compose multiple data source information.

Related genes for the same or a homologous disease tend to have interactions with each other in the PPI network (1). So the protein-protein interaction (PPI) network has become one of the most powerful sources in these studies, and this information has recently been provided in network structures. Integrating the PPI network with other biological data may lead to the discovery of new disease-causing genes. Some studies use local features of the PPI network, such as molecular triangulation (2), shortest path (SP) (3), and direct neighbors (4) to identify candidate genes. Some other studies such as Random Walk with Restart (RWR) (5) and Network Propagation (NP) (6) relay on global network information. The results of local algorithms are more vulnerable because they do not consider indirect functional relationships and so only consider direct interactions (7). On the other hand, global algorithms rely on the overall relationship between the disease gene and the other genes of the PPI network. Although they do not consider genes with poor connections and outliers, they have displayed a better performance than local methods (8).

Biological resources are very prone to containing noise. However, prioritization based on the quality of raw data and the composition of various heterogeneous data sources is helpful to overcome this disadvantage. DIR (9), ToppGene (10), CANDID (11), Endeavor (12) and MetaRanker (13) are examples of these hybrid methods.

In this approach, we propose a novel method called NRSSPrioritize (the acronym for the combination of Network Propagation, Random Walk with Restart, Shortest Paths, and Symptom similarity knowledge for genes Prioritize), which compares disease similarity knowledge with local and global network information to qualify the performance of the prioritization process and overcome the disadvantages of the aforementioned methods. In our method, PPI network genes are scored separately by using NP, RWR and SP methods. Also, we use disease symptom similarity information to score genes. Then, we calculate the final score as a weighted summation of scores. The Multiple Attribute Decision-making (MADM) (14) method was applied to calculate the weight of weighted summation. We applied NRSSPrioritize to Colorectal Cancer (CRC) datasets, that disease being one of the leading causes of death in the world (15, 16). The evaluation proves that using local and global network information in combination with other disease related knowledge leads to more accurate results in comparison to previous methods.

## 2. Material and Methods

### 2.1. Protein-Protein interaction network

We used the PPI network, which represents interaction between proteins. Information regarding to the PPI network is gathered from previous research information (17). Protein interactions in the PPI network are extracted from several databases and each interaction is weighted according to two criteria:

- The interaction was given more weight if it was reported in more studies.
- The weight of interaction that was observed in the large scale experiment is heavier than the other one.

This network is undirected. Our network contains 22997 nodes, which are constructed from 89 subgraphs. Each node represents one protein and the edges are interactions between proteins. There are 10103 isolated nodes in the network that do not have any useful interactional information. So we removed them, and thus 12894 nodes remained. We used the DisGeNET database (18) to extract CRC genes. The DisGeNET database contains 2394 CRC genes. 1121 genes were found in the PPI network and mapped as seed genes.

## 2.2. Genes scores calculation (NRSSPrioritize steps)

We applied NP, RWR, and SP algorithms separately on the PPI network to rank genes. Also, we employed disease similarity information in our scoring. The Dijkstra (19) algorithm was used to calculate the lengths and weight of shortest paths between the seed genes and the other genes. Then we divided the median of the weights of the shortest paths to the seeds to the average of the shortest paths length for each non-seed node. Obviously, a gene with a higher shortest path weight median and a smaller average length is more likely to be associated with disease (20).

*RWR* calculates the proximity between nodes in the network (5). It repeatedly studies the global structure of the PPI network to approximate the relationship rate between nodes. RWR starts with a seed gene. In each step, it is faced with two choices; going to a neighbor of the current node or moving to one of the seed genes. It calculates node scores according to the following equation:

$$y^{t+1} = (1-\propto)Wy^t + \propto y^0 \quad (1)$$

Here $W$ is the adjacency matrix in which its columns are normalized. $y^0$ is a probability vector of the nodes at the first step. We assigned equal values to seed genes, and other gene probabilities are set to zero. $y^t$ represents a probability of all the nodes at step $t$. Also, $\propto$ is a fixed parameter, which is called the restarting probability and it determines the probability of restarting random walk from seed genes ($1-\propto$ is probability of going to a neighbor). In this study, the value of $\propto$ value is 0.15. This algorithm iterates repeats until the difference between $y^t$ and $y^{t+1}$ becomes less than $10^{-6}$ (5).

*NP* uses network flow propagation to prioritize genes. It simulates an information pump that emanates at the seeds. Its idea is very similar to the RWR algorithm with one difference. In the NP algorithm, output and input flows of nodes are both normalized (rows and columns are normalized) (6). After applying these algorithms, three scores are assigned to each non-seed gene.

It is rational to suppose that similar phenotypes have a common causing gene (21, 22), and it has also been proved that the causing gene of diseases with similar symptoms have a shared interacting gene (23). So, we used the Human Symptoms Disease Network (HSDN) (23) to find diseases with similar symptoms to CRC. In this network, nodes represent diseases and the edges are the symptom similarities between them. In the Human Symptoms Disease Network, the subscription rate between diseases and the common MeSH terms between diseases are used to weight the similarity edges. Here, we selected ten top similar diseases to CRC (Fig. 1) and we extracted their causing genes from the DisGeNet database. We calculated the intersection between the causing genes of these top similar diseases and assigned these scores to each non-seed gene. For example, a gene that is associated with one disease, earns a score of one.

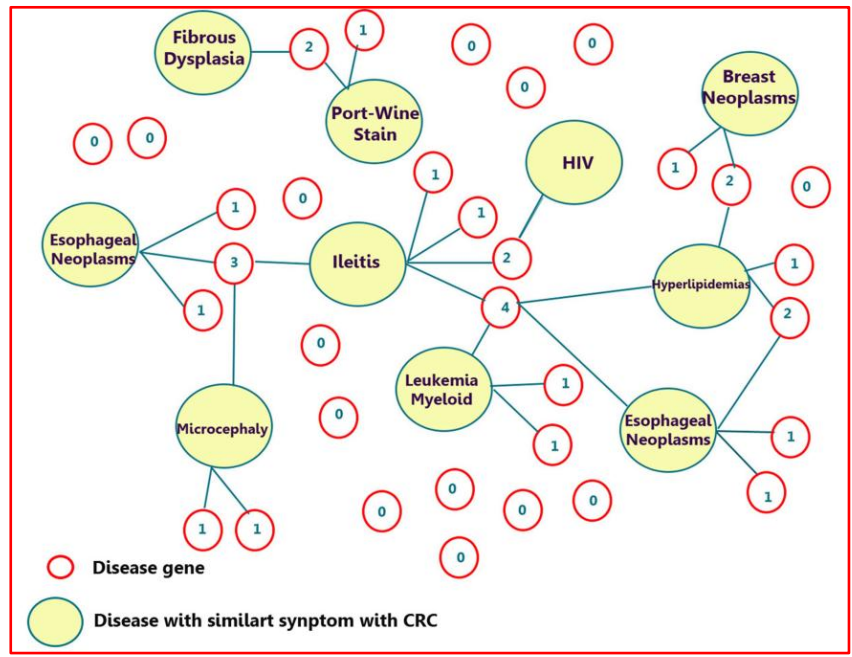

**Figure 1. Top 10 similar diseases to CRC according to their symptoms were obtained from HSDN and their causing genes were extracted from the DisGeNet database. We calculated the intersection between the causing genes of these top similar diseases and assigned this score to each non-seed gene**

**2.3. Combining scores of each gene with MADM**

In the previous steps, the genes received four different scores. Then we used the weighted summation to calculate the final score of each gene. The MADM technique was applied to calculate the weight of each step. In MADM, there are several possible alternatives that are evaluated according to some criteria, and they are then ranked. In this approach, we considered disease symptom similarity, RWR, NP and SP steps as alternatives that are evaluated by leave-one-out cross-validation criteria. In other words, the result of each step was assessed in leave-one-out cross-validation analysis and some criteria were measured according to this analysis to definitively weight each step (Fig. 2).

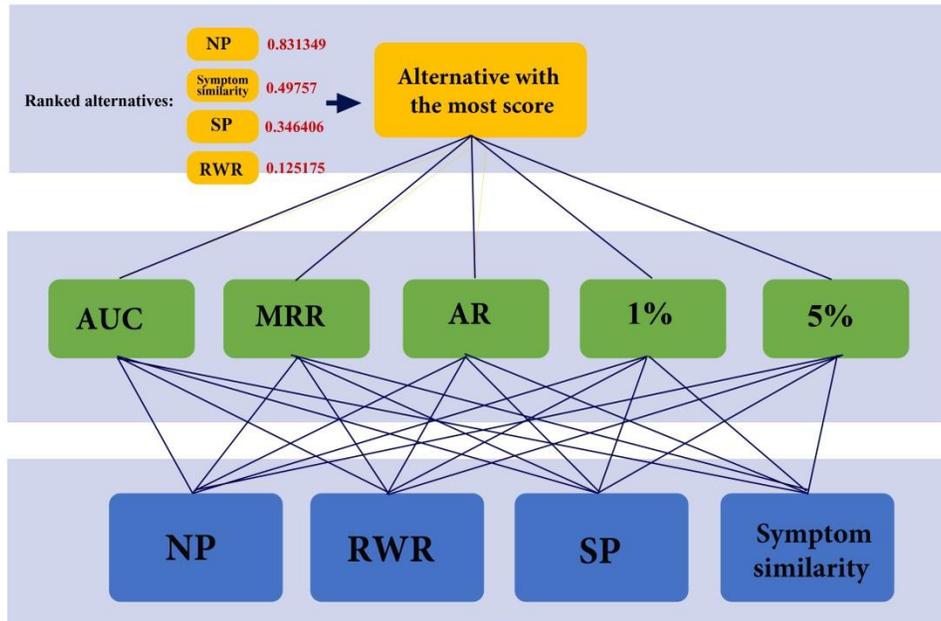

**Figure 2. NP, RWR, SP and symptom similarity scores were calculated for each gene. We used the weighted summation to calculate the final score of each gene. So, MADM was applied to obtain the weight for each step. Four steps are considered as alternatives that are evaluated by criteria. For this reason, the result of each step was assessed in leave-one-out cross-validation analysis and AUC, MRR, AR, 1% and 5% criteria were measured according to this analysis. Finally, a weight was assigned to each step and was then ranked.**

In leave-one-out cross-validation, each seed gene is removed from a seed gene set (target gene) and an artificial linkage interval is constructed on its 99 chromosomal neighbors, which are obtained from the UCSC database (24). These 99 genes and the target gene were then considered to be the candidate set and the remaining seed genes made up the new seed set. We then applied our algorithm to this new candidate set, from which their genes were ranked according to the new seed gene set (Fig. 3).

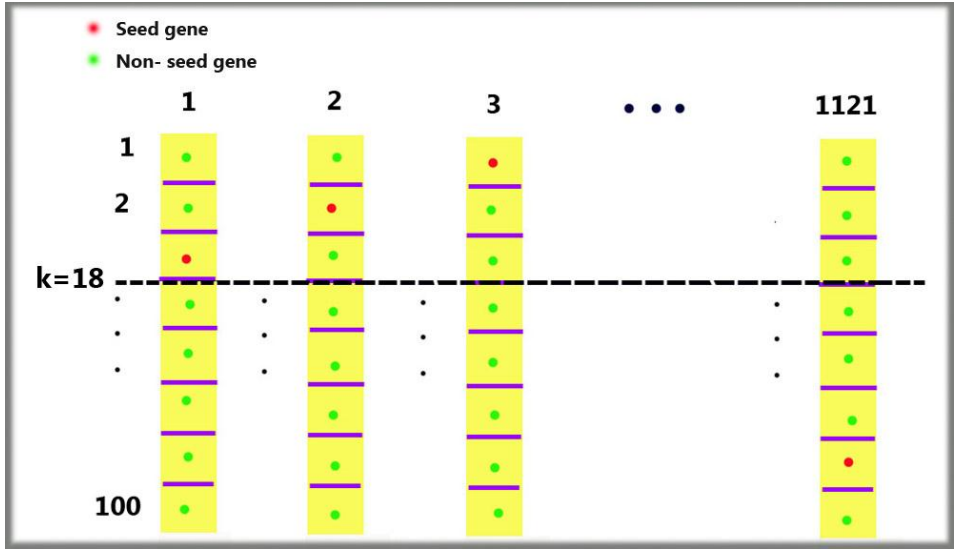

**Figure 3. Leave-one-out cross-validation analysis**

**In each step, one seed gene was removed from the seed genes set (called terget gene) and was considered with 99 chromosomal neighbors as a new candidate set. This new candidate set was ranked using the prioritization method according to the seed genes set information. Finally the target gene position was assessed in a ranked list.**

A Receiver Operating Characteristic (ROC) curve is drawn to show the performance of a specific method with plot sensitivity versus 1-specificity. Also, an Area Under Curve (AUC) measure is calculated for each curve. Here, sensitivity is the percentage of disease genes that are ranked above a specific (k) threshold, and specificity defines the percentage of all genes ranked below the threshold (k). Also, Mean Reciprocal Rank (MRR) is obtained by using the following equation:

$$MRR = \frac{1}{Q}\sum_{i=1}^{Q}\frac{1}{rank_i} \tag{2}$$

Here $rank_i$ is the rank of the target gene in the $i$th candidate set and $Q$ is the count of candidate gene sets. The other measure is an average rank (AR), which is the average rank of the target seeds in their related candidate sets. 1% and 5% are also the other criteria that represent the percentage of target genes in the top 1% and 5% of candidate sets. All these measures are reported for NP, RWR, SP and disease similarity step in Table 1.

After calculating AUC, AR, MRR, 1% and 5% measures, we applied the Analytic Network Process (ANP) (25) method. The ANP considers the relationship and feedback between criteria in the whole system in order to compare them. So the network structure was constructed as you can see in Fig. 4. The following order was utilized to compare the criteria and to weigh them. AUC and MRR globally assess results and thus we gave more importance to them.

$$AUC > MRR > AR > 1\% > 5\%$$

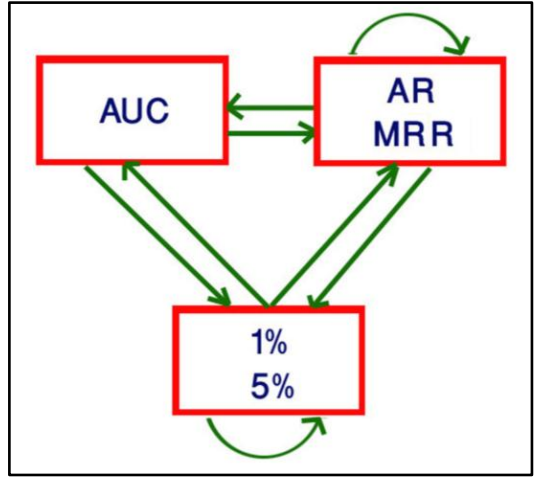

**Figure 4. Network structure for performance criteria in the ANP method.**

Then, these criteria weights was used in the Technique for Order Preference by Similarity to Ideal Solution (TOPSIS) (26) to investigate the RWR, NP, SP and symptom similarity steps (alternatives) according to criteria. The TOPSIS method is based on the concept that the best alternatives should have the shortest distance from the positive-ideal solution and the longest distance from the negative-ideal solution. This means that, positive and negative ideal solutions are built to compare the alternatives against each other and to weigh them. We used these weights in the weighted summation to rank non-seed genes. Notably, other MADM methods are examined and the best result is found using hybrid TOPSIS-ANP, such as analytic hierarchy process (AHP) (27), Fuzzy TOPSIS (FTOPSIS) (28), Fuzzy AHP (FAHP) (29) and FTOPSIS-FAHP methods.

**Table 1. Performance measures for each step of NRSSPrioritize.**

| Methods | AUC | MRR | AR | 1% | 5% |
|---|---|---|---|---|---|
| NP | 0.91 | 0.25 | 9.3 | 0.12 | 0.40 |
| RWR | 0.90 | 0.25 | 10 | 0.12 | 0.40 |
| SP | 0.91 | 0.22 | 9.7 | 0.08 | 0.32 |
| Symptom similarity | 0.88 | 0.34 | 11.88 | 0.18 | 0.48 |

For a more accurate conclusion, we applied the Weighted Discounted Rating System (WDSR) (30) and the N-dimensional Order Statistics (NDOS, used in the Endeavour method) (12) to integrate the outcomes of the four aforementioned steps. These methods compose multiple ranked lists to obtain one rank. We evaluated their performance using leave-one-out cross-validation. Fig. 5 illustrates the ROC curve for each method. As can be seen, weighted summation with TOPSIS-ANP displayed a better performance than WDRS and NDOS by a significant margin.

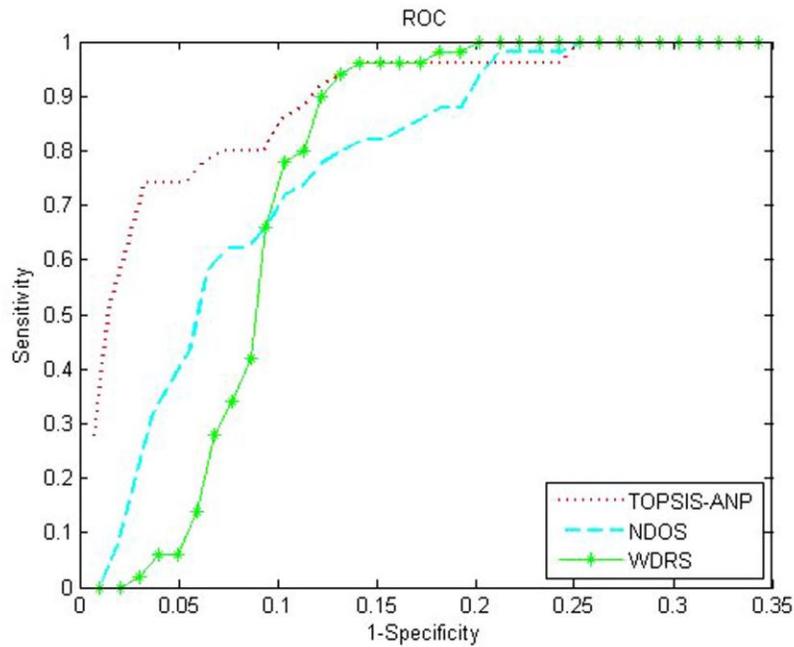

**Figure 5. Performance comparison for Heterogeneous data composition methods**
**Each gene was assigned four scores in our approach and we applied hybrid TOPSIS-ANP to calculate the weight of each score and then obtained the final score by weighted sumation. Also, NDOS and DRS were examined and compared with our result for more reliability.**

## 3. Results and discussion

In this paper, we proposed the NRSSPrioritize method, which integrates protein complex and disease similarity knowledge to utilize the benefits of both. Diseases with similar symptoms probably have common causing genes (21, 22), and also disease related proteins tend to have interactions with each other(1). There are 12894 genes in the PPI network, of which 1121 genes were marked as disease genes (seed genes) and we detected 91% of seed genes (1025 seed genes) in one subgraph of the PPI network. This observation reinforces the hypothesis that disease assosiated genes tend to have interaction with each other in the protein levels. Additionally, we applied the Mann–Whitney U-test to the GSE32323 dataset, which was obtained from the GEO database (31) and observed that 449 of the seeds have significant differential expressions. NP and RWR algorithms utilitize network global information to calculate the proximity between disease genes and other genes, and the SP algorithm relies on local network characteristics. These three methods are separately applied to the PPI network in order to conquer the disadvantage of local and global methods. İt may seem that the simultaneous usage of NP and RWR is useless, but we observed that comparing their scores overcame all performance measures. Finally, all aformentioned information was integrated with weighted summation based on the hybrid TOPSIS-ANP method. NRSSPrioritize was evaluated and compared to other prioritization tools such as Endeavour, ToppGene and DIR, which use multiple data sources to prioritize disease candidate genes. We used ROC analysis to compare various methods (shown in Fig. 6).

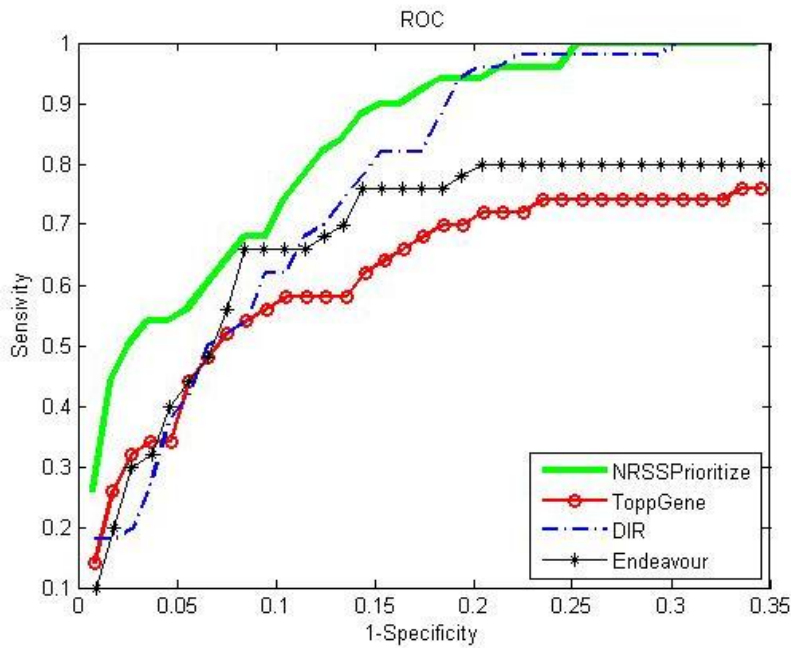

**Figure 6.Comparison of priortization methods**

**A performance comparison has been carried out for NRSSPrioritize, Endeavour, ToppGene and DIR prioritization methods in a leave-one-out cross-validation for CRC causing genes. The figure shows sensitivity versus 1-specificity with diffrent values of k threshold.**

Also, AUC of each curve is calculated for a more accurate comparison. NRSSPrioritize produced the highest AUC with 0.93 and DIR, Endeavour and ToppGene had 0.91, 0.76, and 0.70 AUC values respectively. The MRR for all methods are shown in Fig. 7. Our method achived a 0.42 MRR value, which is higher than the other methods. Additionally in 26% of the cases, NRSSPrioritize ranked the target gene as the first of a candidate gene set and in 54% of the cases, target genes were in a 1 to 5 ranking of the candidate sets. The AR of our method was the lowest amon for all examined algorithms. AR, 1% and 5% measures are shown in Table 2 for all prioritization methods. Overall, all measures showed that our method displayed a better performance in comparison to the other previously examined algorithms.

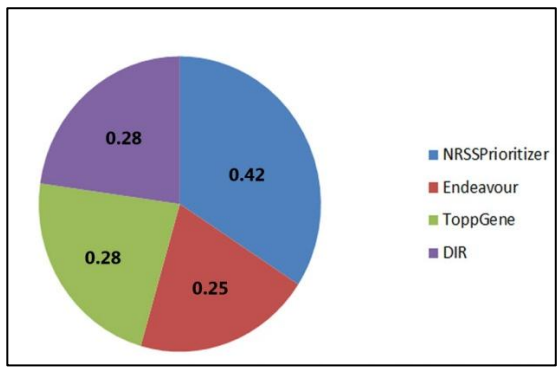

**Figure 7. The MRR value is illustrated for each prioritization method. Our approach 's MMR is the highest.**

## 4. Conclusion

We applied RWR, NP and SP algorithm separately to the PPI network. Then, we scored genes based on disease similarity knowledge. We combined these four scores by weighted summation. We used the TOPSIS-ANP MADM method to calculate the weight of each score. With this strategy, we combined protein complexity information with specialized disease knowledge. Also, NP, SP and RWR are network-based methods and using them at same time led to decreasing the noise of the PPI network. Our achievements prove that considering various characteristics of genes is important in disease gene identification and that we can overcome the disadvantages of existing methods by combining different types of knowledge.

**Table 2. AR, 1% and 5% measure values for our method and other prioritization methods.**

| Methods | AR | 1% | 5% |
|---|---|---|---|
| NRSSPrioritizer | 7.1 | 0.26 | 0.54 |
| Endeavour | 10.5 | 0.10 | 0.40 |
| TopGene | 10.94 | 0.14 | 0.34 |
| DIR | 9.6 | 0.18 | 0.38 |

**Acknowledgments**

This research was supported financially by a MSc grant from research affairs of University of Tehran.

**References**
1       Gandhi, T.K., Zhong, J., Mathivanan, S., Karthick, L., Chandrika, K.N., Mohan, S.S., Sharma, S., Pinkert, S., Nagaraju, S., Periaswamy, B. *et al.* (2006) Analysis of the human protein interactome and comparison with yeast, worm and fly interaction datasets. *Nat Genet*, **38**, 285-293.
2       Krauthammer, M., Kaufmann, C.A., Gilliam, T.C. and Rzhetsky, A. (2004) Molecular triangulation: bridging linkage and molecular-network information for identifying candidate genes in Alzheimer's disease. *PNAS*, **101**, 15148-15153.
wrtest
3       George, R.A., Liu, J.Y., Feng, L.L., Bryson-Richardson, R.J., Fatkin, D. and Wouters, M.A. (2006) Analysis of protein sequence and interaction data for candidate disease gene prediction. *Nucl Acids Res*, **34**, 691-698.
4       Oti, M., Snel, B., Huynen, M.A. and Brunner, H.G. (2006) Predicting disease genes using protein-protein interactions. *J Med Genet* **43**, 691-698.
5       Kohler, S., Bauer, S., Horn, D. and Robinson, P.N. (2008) Walking the Interactome for Prioritization of Candidate Disease Genes. *Am J Hum Genet* **82**, 949-958.
6       Vanunu, O., Magger, O., Ruppin, E., Shlomi, T. and Sharan, R. (2010) Associating Genes and Protein Complexes with Disease via Network Propagation. *PLoS Comput Biol*, **6**, e1000641.
7       Erten, S., Bebek, G., Ewing, R.M. and Koyutürk, M. (2011) DADA: Degree-Aware Algorithms for Network-Based Disease Gene Prioritization. *BIODATA MIN*, **4**, 1-20.
8       Doncheva, N.T., Kacprowski, T. and Albrecht, M. (2012) Recent approaches to the prioritization of candidate disease genes. *Wiley Interdiscip Rev Syst Biol Med*, **4**, 429-442.
9       Chen, Y., Wang, W., Zhou, Y., Shields, R., Chanda, S.K., Elston, R.C. and Li, J. (2011) In Silico Gene Prioritization by Integrating Multiple Data Sources. *PLoS ONE*, **6**, e21137.
10      Chen, J., Xu, H., Aronow, B.J. and Jegga, A.G. (2007) Improved human disease candidate gene prioritization using mouse phenotype. *BMC Bioinform.*, **8**, 1-13.
11      K. A. Hutz JE and HL, M. (2008) CANDID: a flexible method for prioritization candidate genes for complex human traits. *Genet Epidemiol*, in press., 779-811.
12      Aerts, S., Lambrechts, D., Maity, S., Loo, P.V., Coessens, B., Smet, F.D., Tranchevent, L.C., Moor, B.D., Marynen, P., Hassan, B. *et al.* (2006) Gene prioritization through genomic data fusion. *Nat Biotechnol*, **24**, 537-544.
13      Pers, T.H., Hansen, N.T., Lage, K., Koefoed, P., Dworzynski, P., Miller, M.L., Flint, T.J., Mellerup, E., Dam, H., Andreassen, O.A. *et al.* (2011) Meta-analysis of heterogeneous data sources for genome-scale identification of risk genes in complex phenotypes. *Genet Epidemiol*, **35**, 318-332.
14      Figueira, J., Greco, S. and Ehrgott, M. (2005) *Multiple Criteria Decision Analysis: State of the Art Surveys Series*.
15      Marmot, M., Atinmo, T., Byers, T., Chen, J., Hirohata, T., Jackson, A., James, W., Kolonel, L., Kumanyika, S. and Leitzmann, C. ( 2007 Food, nutrition, physical activity, and the prevention of cancer: a global perspective . AICR, in press.
16      Parkin, D.M., Whelan, S.L., Ferlay, j., Teppo, L. and Thomas, D.B. (2002), In *The World Health Organization and The International Agency for Research on Cancer*. IARC Scientific, Lyon, in press.
17      Lage, K., Karlberg, E.O., Storling, Z.M., Olason, P.I., Pedersen, A.G., Rigina, O., Hinsby, A.M., Tumer, Z., Pociot, F., Tommerup, N. *et al.* (2007) A human phenome-interactome network of protein complexes implicated in genetic disorders. *Nat Biotechnol*, **25**, 309-316.

18	Pinero, J., Q. Rosinach, N., Bravo, A., D. Pons, J., B. Mehren, A., Baron, M., Sanz, F. and Furlong, L.I. (2015) DisGeNET: a discovery platform for the dynamical exploration of human diseases and their genes. *Database (Oxford)*, **2015**, bav028.
19	Dijkstra, E.W. (1959) a note on two pleoblem inconnexion with graphs. *Numer. Math*, in press.
20	George, R.A., Liu, J.Y., Feng, L.L., Bryson-Richardson, R.J., Fatkin, D. and Wouters, M.A. (2006) Analysis of protein sequence and interaction data for candidate disease gene prediction. *Nucleic acids research*, **34**, 691-698.
21	Oti, M. and Brunner, H.G. (2007) The modular nature of genetic diseases. *Clin Genet*, **71**, 1-11.
22	Limviphuvadh, V., Tanaka, S., Goto, S., Ueda, K. and Kanehisa, M. (2007) The commonality of protein interaction networks determined in neurodegenerative disorders (NDDs). *Bioinformatics (Oxford, England)*, **23**, 2129-2138.
23	Zhou, X., Menche, J., Barabasi, A.L. and Sharma, A. (2014) Human symptoms-disease network. *Nat Commun*, **5**, 4212.
24	Dreszer, T.R., Karolchik, D., Zweig, A.S., Hinrichs, A.S., Raney, B.J., Kuhn, R.M., Meyer, L.R., Wong, M., Sloan, C.A., Rosenbloom, K.R. *et al.* (2012) The UCSC Genome Browser database: extensions and updates 2011. *Nucleic Acids Res*, **40**, D918-D923.
25	Saaty, T.L. (1996), In *Decision Making with Dependence and Feedback*. RWS Publications, Pennsylvania, in press.
26	Yoon, K. (1987) A reconciliation among discrete compromise situations. *J. Oper. Res. Soc.*, **38**, 277–286.
27	Saaty, T.L. and Peniwati, K. (2013) *Group decision making: drawing out and reconciling differences*. RWS publications.
28	Lu, J., Zhang, G. and Ruan, D. (2007) *Multi-objective group decision making: methods, software and applications with fuzzy set techniques*. Imperial College Press.
29	Gwo-Hshiung, T. (2010) *Multiple attribute decision making: methods and applications*. CRC press.
30	Li, Y. and Patra, J.C. (2010) Integration of multiple data sources to prioritize candidate genes using discounted rating system. *BMC Bioinform.*, **11**, 1-10.
31	Barrett, T., Troup, D.B., Wilhite, S.E., Ledoux, P., Rudnev, D., Evangelista, C., Kim, I.F., Soboleva, A., Tomashevsky, M., Marshall, K.A. *et al.* (2009) NCBI GEO: archive for high-throughput functional genomic data. *Nucleic Acids Res*, **37**, D885-D890.